\documentstyle[preprint,aps,epsfig,eqsecnum,tighten]{revtex} 
\tightenlines

\newcommand{\nc}{\newcommand}
\nc{\beq}{\begin{equation}}   \nc{\eeq}{\end{equation}}
\nc{\bea}{\begin{eqnarray}}   \nc{\eea}{\end{eqnarray}}
\nc{\baa}{\begin{array}}      \nc{\eaa}{\end{array}}
\nc{\bit}{\begin{itemize}}    \nc{\eit}{\end{itemize}}
\nc{\ben}{\begin{enumerate}}  \nc{\een}{\end{enumerate}}
\nc{\bce}{\begin{center}}     \nc{\ece}{\end{center}}
\def\beqa{\begin{eqnarray}}
\def\eeqa{\end{eqnarray}}
\def\be{\beq}
\def\ee{\eeq}
\def\to{\rightarrow}

\def\half{{1\over 2}}

\def\lsim{\mathrel{\raise.3ex\hbox{$<$\kern-.75em\lower1ex\hbox{$\sim$}}}}
\def\gsim{\mathrel{\raise.3ex\hbox{$>$\kern-.75em\lower1ex\hbox{$\sim$}}}}
\def\ie{{\it i.e.\ }}
\def\eg{{\it e.g.\ }}

\begin{document}

\draft
\tightenlines

\topmargin = -1.0cm
\overfullrule 0pt
%\preprint{\vbox{\hbox{IFUSP-DFN/01-060}
%\hbox{hep-ph/0105196}
%}}
\title{
Probing Flavor Changing Neutrino Interactions Using Neutrino Beams 
from a Muon Storage Ring} 
\author{ 
A. M. Gago$^{1,2}$~\footnote{
E-mail: agago@charme.if.usp.br\\
$^{\ \ \dagger}$E-mail: guzzo@ifi.unicamp.br\\
$^{\ \ \ddagger}$E-mail: nunokawa@ifi.unicamp.br\\
$^{\ \ \natural}$E-mail: teves@charme.if.usp.br \\
$^{\ \ \sharp}$E-mail: zukanov@charme.if.usp.br}
M.  M. Guzzo$^{3}$$^{\ \dagger}$,
H. Nunokawa$^{3}$$^{\ \ddagger}$ 
W. J. C. Teves$^{1}$$^{\ \natural}$\\
and 
R. Zukanovich Funchal$^{1}$$^{\ \sharp}$
}
\address{\sl 
$^1$ Instituto de F\'{\i}sica, Universidade de S\~ao Paulo 
C.\ P.\ 66.318, 05315-970 S\~ao Paulo, Brazil\\
$^2$  Secci\'{o}n F\'{\i}sica, Departamento de Ciencias, Pontificia Universidad Cat\'olica del Per\'u \\
Apartado 1761, Lima, Per\'u   \\
$^3$ Instituto de F\' {\i}sica Gleb Wataghin, Universidade Estadual de Campinas, UNICAMP\\    
13083-970 -- Campinas, Brazil.}

\maketitle
\vspace{-0.4cm}
%%%%%%%%%%%%%%%%%%%%%%%%%%%%%%%%%%%%%%%%%%%%%%%%%%%%%%%%%%%%%%%%%%%
\hfuzz=25pt
\begin{abstract}
We discuss the capabilities of a future neutrino factory 
based on intense neutrino beams from a muon storage ring
to explore the non-standard neutrino matter interactions, 
which are assumed to be sub-leading effects in the standard 
mass induced neutrino oscillations. 
The conjunction of these two mechanisms will magnify
fake $CP$ violating effect in the presence of matter
which is not coming from the $CP$ phase in the neutrino
mixing matrix. We show that such fake $CP$ violation can
be observed in neutrino factory experiments by measuring 
the difference between the neutrino and anti-neutrino 
probabilities.
In order to perform such test, we consider three neutrino 
flavors, admitting  the mixing parameters in the range 
consistent with the oscillation solution to the atmospheric and 
the solar neutrino problems, as well as with 
the constraints imposed by the reactor neutrino data. 
We show that with a 10 kt detector with 5 years of operation, 
a stored muon energy $E_\mu \ge 20$ GeV, $2 \times 10^{20}$ 
muon decays per year, and a baseline $L \sim 732$ km, 
such a neutrino facility can probe the non-standard flavor 
changing neutrino interactions down to the level of 
($10^{-3}-10^{-2})~G_F$, in both 
$\nu_\mu \to \nu_\tau/ \bar \nu_\mu \to \bar \nu_\tau$
and $\nu_e \to \nu_\tau/\bar \nu_e \to \bar \nu_\tau$ 
modes.

\end{abstract}
%\vskip2pc]
\newpage

%%%%%%%%%%%%%%%%%%%%%%%%%%%%%%%%%%%%%%%%%%%%%%%%%%%%%%%%%%%%%%%%%%%%%%
\section{Introduction}
\label{sec:intro}
%%%%%%%%%%%%%%%%%%%%%%%%%%%%%%%%%%%%%%%%%%%%%%%%%%%%%%%%%%%%%%%%%%%%%%

The atmospheric neutrino experiments~\cite{atmexp} 
present today compelling evidence in favor of $\nu_\mu$ disappearance. 
The solar neutrino experiments~\cite{solarexp} as well, are 
strongly indicating $\nu_e$ disappearance. 
The Los Alamos Liquid Scintillator Neutrino Detector (LSND)~\cite{lsnd}
has also reported results consistent with neutrino flavor conversion. 

The most plausible mechanism of such neutrino flavor 
disappearance or conversion 
is the oscillation of neutrino 
due to quantum interference driven by neutrino mass squared 
differences $\Delta m^2$~\cite{Pontecorvo,MNS}, 
which can happen when neutrino flavor eigenstates 
are supposed to be coherent superpositions of 
neutrino mass eigenstates~\cite{MNS}. 
Since the impressive results from 
the LEP experiments~\cite{pdg00,lep}, 
supported recently by the direct observation of 
$\nu_\tau$ by DONUT~\cite{donut}, established that there are at least 
three active neutrino flavors, it is unavoidable to try to 
understand such mass induced neutrino oscillations in a full three 
generation scenario. 

Under the assumption of such mass induced neutrino oscillation
in three flavor framework, 
atmospheric neutrino data~\cite{sk-tau} can be explained by pure 
$\nu_\mu \to \nu_\tau$ oscillations in vacuum~\cite{atmValencia}, 
though the possibility of having contributions from non-negligible 
$\nu_\mu \to \nu_e$ oscillations is still not 
discarded~\cite{atm3nu}, even after taking into account 
the constraints coming from the CHOOZ reactor 
experiment~\cite{chooz}. 
On the other hand, solutions to the solar neutrino problem 
can be provided either by the matter enhanced resonant neutrino conversion, 
the MSW effect~\cite{MSW,msw-sol}, or by vacuum oscillations with a 
typical wavelength of the order of the Sun-Earth distance~\cite{justso}
with the pure two generation mixing or with small additional 
participation of a third neutrino on top of the two generation mixing. 
In this work, we do not take into account signals of 
the LSND experiment, since simultaneous explanations 
of the LSND data together with atmospheric as well 
as solar neutrinos data require the presence of a fourth sterile 
(electroweak singlet) neutrino, 
which is beyond the scope of this paper. 

Several non-standard (or exotic) explanations of atmospheric 
as well as solar neutrino data, which do not 
necessarily invoke neutrino mass and/or mixing, also have been 
suggested.
For atmospheric neutrinos, $\nu_\mu$ conversion or disappearance 
mechanisms induced by neutrino decay~\cite{pakvasa},  
flavor changing neutrino interactions~\cite{us} and 
quantum decoherence~\cite{atm_deco} have been proposed 
to explain these data. 
Among them, it was found that the flavor changing solution can 
not explain well the SK upward going muon data~\cite{fc_upgo}
and also it is disfavored by the results coming from K2K~\cite{K2K}. 
For solar neutrinos, there are also several 
proposals to explain the data by 
invoking non-standard neutrino properties such 
as magnetic moment~\cite{rsfp}, 
non-standard neutrino interactions~\cite{fc-sol} 
or even by a tiny violation of the equivalence principle~\cite{vep}. 

While some of these non-standard solutions can explain 
quite well the data, it is generally believed that the most plausible 
solutions are provided by the standard mass induced oscillation 
mechanism, because mass and mixing are the simplest
extension of the standard model and moreover, 
it is the unique mechanism which can explain 
well both atmospheric and solar neutrino observations 
at the same time, without additional neutrino properties. 

However, even in this case, we consider that it is 
quite possible that some non-standard neutrino properties
could induce some sub-leading effect in addition to the 
the standard mass induced neutrino oscillation without 
causing any inconsistency with the present observations. 
We are particularly interested in such sub-leading effects  
induced by flavor changing neutrino interactions. These type of 
interactions were first suggested in Ref.~\cite{wol},   
and then discussed as a possible mechanism to solve the solar neutrino 
problem in Ref.~\cite{GMP,roulet}.

Indeed, many models of neutrino mass~\cite{see-saw,radmod,mssm} are 
also a natural source of non-standard neutrino matter interactions (NSNI) 
which can be, phenomenologically, classified into two types :  
flavor changing neutrino interactions (FCNI) and 
flavor diagonal neutrino interactions (FDNI), 
as we will discuss in detail in our work. 

Therefore, we believe that even if mass induced oscillations really 
take place in nature and such NSNI are not playing any relevant roles 
in explaining the observed neutrino data, it could be important to study 
NSNI interactions as secondary contributions, in order to gain some 
handle on new physics beyond the electroweak standard model. 
The crucial point of our idea is to use the fact that  
the simultaneous presence of neutrino masses and NSNI can enhance the 
difference in the conversion probabilities between neutrino and 
anti-neutrino channels in matter causing fake 
$CP$ violation~\cite{nufact00}. 

Our aim here is to test such interactions in a future neutrino factory. 
A neutrino factory~\cite{nufactory}, which could be a milestone 
for a future muon-collider project, presents a possibility to go 
beyond the goal of measuring with great precision 
the oscillation parameters. 
It will offer us the chance to measure separately 
the oscillating probabilities in $CP$-conjugate channels, such as 
$\nu_\mu \to \nu_e,\nu_\tau$ and $\bar \nu_\mu \to \bar \nu_e,\bar \nu_\tau$, 
with high intensity neutrino beams produced 
by a muon storage ring~\cite{nufactory}, 
after neutrinos travel through several hundreds of kilometers in 
the Earth mantle before reaching the detector. 

There are already quite a few very interesting studies that have shown 
the prospects of neutrino factories to measure the mixing parameters and 
obtaining the pattern of neutrino masses~\cite{nufactory,flpr,bgrw}, 
to test matter effects~\cite{flpr,bgrw}, $CP$ violation in the leptonic 
sector~\cite{romanino,gpz}, $CPT$-odd interactions~\cite{bpww} 
and distinguish the mass induced solution for the atmospheric 
neutrino anomaly from the decoherence one~\cite{gstz}, 
or possible neutrino oscillation induced by 
violation of the equivalence principle~\cite{Datta_vep}. 
Signals of $R-$parity violating supersymmetric interactions which can be 
mistaken for neutrino oscillations in neutrino factory experiments were 
also investigated in Ref.~\cite{datta-gandhi}, where it is found that 
these novel contributions basically do not affect the $\tau$ event rate for  
baselines greater than 200 km, except if the new couplings are close to their 
perturbative limit.

Here, we investigate the capacity of neutrino factory experiments  
to probe neutrino oscillations beyond the standard mechanism 
assuming mixing parameters compatible with atmospheric, solar 
and reactor neutrino experiments and by so doing to discover 
new physics or put restrictive limits on NSNI models.  
Our results will be presented as a function of the 
distance between source and detector for two particular muon energies. 

In Sec. \ref{sec:form}, we define the formalism that will be used throughout  
this paper. In Sec.\ \ref{sec:setup}, we describe the various  experimental 
set-ups we will consider, as well as the details of our experimental 
simulations. In Sec.\ \ref{sec:res}, we show our predictions for the 
$\nu_\mu \to \nu_\tau$ and $\nu_e \to \nu_\tau$ channels  
with and without $CP$ violation. Finally in Sec. \ref{sec:conc}, we 
discuss our results and present our conclusions.

%%%%%%%%%%%%%%%%%%%%%%%%%%%%%%%%%%%%%%%%%%%%%%%%%%%%%%%%%%%%%%%%%%%%%%
\section{Modified Oscillation Formalism with NSNI}
\label{sec:form}
%%%%%%%%%%%%%%%%%%%%%%%%%%%%%%%%%%%%%%%%%%%%%%%%%%%%%%%%%%%%%%%%%%%%%%

There are currently several experimental bounds on NSNI coming 
from the non-observation of lepton flavor violating 
process~\cite{pdg00}, from violation of lepton flavor 
universality, or from neutrino scattering data. 
Model independent analysis of these experimental constraints
derived from $SU(2)_L$ related interactions, that can be found 
in Refs.\ \cite{bgp,fc-sol}, in general, gives 
stronger bound for $\nu_e-\nu_\mu$ channel, 
typically set the maximal strength of NSNI normalized by 
the electro weak interactions strength, i.e. $G_F$, 
to be at the level much smaller than one-percent 
whereas the bounds for $\nu_\mu-\nu_\tau$ and 
$\nu_e-\nu_\tau$ channels are substantially weaker 
typically at the level of $\sim 10^{-2} - 10^{-1}$ 
in unit of $G_F$. 
Therefore, in this work, we will only consider the
two channels $\nu_\mu-\nu_\tau$, $\nu_e-\nu_\tau$ 
and their anti-neutrino ones, which are less 
constrained from the laboratory experiments. 

In the usual three generation neutrino oscillation framework, we can 
define the correspondence between  neutrino mass eigenstates and neutrino 
interaction eigenstates by,

\be
\left[ 
\begin{array}{c} 
\nu_e \\
\nu_\mu \\
\nu_\tau
\end{array}
\right]
= U 
\left[ 
\begin{array}{c} 
\nu_1 \\
\nu_2 \\
\nu_3
\end{array}
\right],
\label{rel}
\ee
where $\nu_i$ ($i=1,2,3$) are the neutrino mass eigenstates, 
and $\nu_\alpha$ ($\alpha=e,\mu,\tau$) are the neutrino flavor 
eigenstates with the standard parameterization for 
Maki-Nakagawa-Sakata~\cite{MNS} mixing matrix $U$ given by, 
\begin{equation}
U=\left[
\begin{array}{ccc}
c_{12}c_{13} & s_{12}c_{13} & s_{13}e^{-i\delta } \\
-s_{12}c_{23}-c_{12}s_{23}s_{13}e^{i\delta } &
c_{12}c_{23}-s_{12}s_{23}s_{13}e^{i\delta } & s_{23}c_{13} \\
s_{12}s_{23}-c_{12}c_{23}s_{13}e^{i\delta } &
-c_{12}s_{23}-s_{12}c_{23}s_{13}e^{i\delta _{}} & c_{23}c_{13}
\end{array}
\right],  \label{km}
\end{equation}  
where $c_{ij}$ and $s_{ij}$ denote the cosine and the sine of $\theta_{ij}$  
and $\delta$ is the $CP$ violating phase.

\subsection{NSNI in $\nu_\mu \to \nu_\tau$ channel}
\label{nmutau}

If we consider the possibility of NSNI only in the $\nu_\mu \to \nu_\tau$ 
channel, the evolution Hamiltonian in matter has the form,
\be   
i{\displaystyle\frac{d}{dr}}  \left( \begin{array}{c} \nu_e \\ \nu_\mu \\ \nu_\tau \end{array} \right) 
=  \left[ U \left(\begin{array}{ccc} 0 & 0 & 0 \\
0 & \Delta_{21} & 0\\
0 & 0 & \Delta_{31}
\end{array} \right) U^\dagger +
\left( \begin{array}{ccc} V_e(r)  & 0 & 0 \\
0 & 0 &  \epsilon^f_{\mu\tau} V_f(r)\\ 
0 & \epsilon^f_{\mu\tau} V_f(r)& \epsilon '^f_{\mu\tau} V_f(r)
\end{array} \right) \right]
\left( \begin{array}{c} \nu_e \\ \nu_\mu \\ \nu_\tau  \end{array} \right),
\label{fc1} 
\ee
where $\Delta_{ij}=\Delta m^2_{ij}/2E=(m^2_i-m^2_j)/2E, \, i,j=1,2,3$, 
$E$ is the neutrino energy, $V_f(r)= \sqrt{2} \,G_F n_f(r), f=u,d,e$, 
$V_f(r) \epsilon^f_{\mu\tau}$ is the flavor-changing 
$\nu_\mu + f \to \nu_\tau +f$ forward scattering amplitude 
with the interacting fermion $f$ (electron, $d$ or $u$ 
quark) and $V_f(r) \epsilon_{\mu\tau} '^f$ is the difference between the 
flavor diagonal $\nu_\mu - f$ and $\nu_\tau - f$ elastic forward scattering
amplitudes, with $n_f(r)$ being the number density of the fermions 
which induce such processes. 
In the evolution Eq. (\ref{fc1}),  
$\epsilon^f_{\mu\tau}$ and $\epsilon '^f_{\mu\tau}$ 
are the phenomenological parameters which characterize 
the strength of FCNI and FDNI, respectively. 
The fermion number density $n_f(r)$ can be written in terms of 
the matter density $\rho$ as $n_f(r)=\rho(r) Y_f$, where $Y_f$ is the 
fraction of the fermion $f$ per nucleon, $\sim 1/2$ for electrons 
and $\sim 3/2$ for $u$ or $d$ quarks.  
In all cases, electron neutrinos will coherently scatter off the electrons 
present in matter through the  standard electroweak charged currents which 
introduce non trivial contributions to the neutrino evolution equations. 
These contributions are taken into account by the diagonal term  
$V_e(r)$ in Eq.~(\ref{fc1}). 

A similar form applies to the evolution of anti-neutrinos except that  
$V_{e,f}(r) \to - V_{e,f}(r)$ and $U \to U^*$.  
We assume the density profile of the Earth to be the one given by the 
Preliminary Reference Earth Model~\cite{dens} and solve these 
evolution equations numerically to compute 
the oscillation probabilities $P(\nu_\mu \to \nu_\tau)$ and 
$P(\bar \nu_\mu \to \bar \nu_\tau)$.

\subsection{NSNI in $\nu_e \to \nu_\tau$ channel}
\label{nentau}

Now, if we consider NSNI only in the $\nu_e \to \nu_\tau$ 
channel, the evolution Hamiltonian in matter will have the form,
\be   
i{\displaystyle\frac{d}{dr}} \left( \begin{array}{c} \nu_e \\ \nu_\mu \\ \nu_\tau \end{array} \right) 
=  \left[ U \left(\begin{array}{ccc} 0 & 0 & 0 \\
0 & \Delta_{21} & 0\\
0 & 0 & \Delta_{31}
\end{array} \right) U^\dagger +
\left( \begin{array}{ccc} V_e(r)  & 0 & \epsilon^f_{e\tau} V_f(r) \\
0 & 0 & 0\\
\epsilon^f_{e\tau} V_f(r) & 0 &  \epsilon'^f_{e\tau} V_f(r)
\end{array} \right) \right]
\left( \begin{array}{c} \nu_e \\ \nu_\mu \\ \nu_\tau  \end{array} \right).
\label{fc2} 
\ee
Again for anti-neutrinos we have to make the following substitution 
$V_{e,f}(r) \to - V_{e,f}(r)$ and $U \to U^*$. We obtain the 
oscillation probabilities $P(\nu_e \to \nu_\tau)$ and 
$P(\bar \nu_e \to \bar \nu_\tau)$ by numerically solving Eq.~(\ref{fc2}) 
for neutrinos and the corresponding modified one for anti-neutrinos.

In the following, we will drop the $f$ superscript of $\epsilon$, 
$\epsilon^\prime$ and consider interactions only with either $u$ or $d$ 
quarks, since limits on interactions with electrons can be obtained by  
 a simple re-scale of our plots. 
If NSNI are caused by electrons, $\epsilon$ and $\epsilon^\prime$ 
parameters must be increased by a factor 3 to get the same effect
presented in this paper. 

\subsection{Useful Two Generation Formulae}
\label{useful}

It is instructive to write down the analytical expression for the conversion 
probability in two generation considering constant matter density. This 
will help to understand the full-fledged three generation numerical results.
The probability of $\nu_\alpha \to \nu_\tau$, $\alpha=e, \mu$, 
conversion can be written as
\bea
P(\nu_\alpha \to \nu_\tau)& = 
& \displaystyle \frac{4 \left (\displaystyle \frac{\Delta}{2}
\sin 2\theta + \epsilon_{\alpha \tau} V_f \right )^2}
{4 (\displaystyle 
\frac{\Delta}{2}\sin 2\theta + \epsilon_{\alpha \tau} V_f )^2 
+ (\Delta \cos 2\theta + \epsilon^\prime_{\alpha \tau} 
V_f- \delta_{e \alpha} V_e)^2} \nonumber \\
& & \displaystyle \times \sin^2 \left \{ \frac{L}{2}
\left [ 4 (\frac{\Delta}{2}\sin 2\theta + \epsilon_{\alpha \tau} V_f)^2+ (\Delta\cos 2\theta + \epsilon^\prime_{\alpha \tau} V_f- \delta_{e \alpha} V_e)^2 \right ]^\half \right \},
\label{prob2g}
\eea
where $\Delta$ and $\sin 2 \theta$ will be chosen depending on the 
oscillation mode.
The probability of $\bar \nu_\alpha \to \bar \nu_\tau$ 
conversion is analogous to the above except that 
$V_{e,f} \to - V_{e,f}$. 

One can easily see that the probability is much more sensitive 
to $\epsilon_{\alpha \tau}$ than to $\epsilon^\prime_{\alpha \tau}$, 
since $\epsilon_{\alpha \tau}$ appears in the numerator 
and in the denominator of the amplitude in  Eq.~(\ref{prob2g}), 
while $\epsilon^\prime_{\alpha \tau}$ appears only in the denominator.
Moreover, the terms involving $\epsilon_{\alpha \tau}$ are increased by 
a factor 4 with respect to the terms involving $\epsilon^\prime_{\alpha \tau}$.
These features become more prominent if the distance is not
so large and the contribution from FCNI are small as we will 
see below. 

Let us take some typical values for the mixing parameters and the neutrino 
energy we are going to consider in this work, 
$\Delta m^2\sim \text{few} \times 10^{-3}$ eV$^2$, 
$E$ = few $\times 10$ GeV and 
not very large distance, $L \lsim 1000$ km.  
In this case, taking the NSNI effects to be small, \ie, 
$\epsilon, \epsilon' \ll 1$, 
the probability in Eq.~(\ref{prob2g}) can be
approximated,  up to the first order in the NSNI parameters, 
as follows, 
\begin{equation}
P(\nu_\alpha \to \nu_\tau)
\simeq \frac{1}{4}\Delta \sin 2\theta
\displaystyle 
\left(\Delta \sin 2\theta + 4~\epsilon_{\alpha \tau} V_f \right)
L^2. 
\label{prob2g_approx}
\end{equation}
Note that there is no dependence on the  
$\epsilon'_{\alpha \tau}$ parameter in the above  probability.  
For the corresponding anti-neutrino channel, 
$V_f$ must be replaced by $-V_f$ in 
Eq. (\ref{prob2g_approx}), which lead to a different
probability. 
It is important  to point out that, as we can see 
from Eq. (\ref{prob2g}), only the presence of 
NSNI term alone without neutrino mass ($\Delta \to 0$) 
can not cause the neutrino and anti-neutrino 
probabilities to differ. 

In order to extract the information on NSNI, 
without knowing the very precise values of mixing parameters, 
we must compare in some way, 
the probability for neutrinos and anti-neutrinos. 
In this work, we consider the ratio of the 
expected number of events for anti-neutrinos and neutrinos, which can be 
approximately inferred, for small values of $\epsilon_{\alpha \tau}$  
taking into account the difference in the 
detection cross sections ($\sigma_{\bar\nu_\tau}$ and $\sigma_{\nu_\tau}$),
from the simple expression,

\begin{eqnarray}
R(P)
& \equiv & 
\frac{ \sigma_{\bar \nu_\tau} P(\bar{\nu}_\alpha \to \bar{\nu}_\tau) }
{\sigma_{\nu_\tau} P(\nu_\alpha \to \nu_\tau)}
\simeq \half -
\frac{4~\epsilon_{\alpha \tau} V_f }{ \Delta \sin 2\theta+ 
4~\epsilon_{\alpha \tau} V_f}.
\label{ratio_approx}
\end{eqnarray}
{}From the expressions in 
Eqs.~(\ref{prob2g_approx}) and 
(\ref{ratio_approx}), it is clear that the 
oscillation probabilities are much more sensitive to
the FCNI parameter, $\epsilon_{\alpha \tau}$, than to the FDNI parameter, 
$\epsilon'_{\alpha \tau}$,   
and moreover, in the ratio, the FCNI effect 
does not depend on the distance as long as it is not very 
large, \ie $L\lsim 1000$ km, 
this will be confirmed in the following sections. 

In this work, we assume that $\Delta m^2_{31}$ is 
positive and do not consider the negative case 
(inverted hierarchy) because such case will covered 
as we consider both the positive and negative sign of 
$\epsilon_{\alpha\tau}$. 
>From Eq.~(\ref{ratio_approx}) we can see that 
the same effect is obtained when 
$ (\Delta m^2_{31}, \epsilon_{\alpha\tau})$ 
is replaced by  
$(-\Delta m^2_{31}, -\epsilon_{\alpha\tau})$, 
which is valid as long as $L\lsim 1000$ km.

%%%%%%%%%%%%%%%%%%%%%%%%%%%%%%%%%%%%%%%%%%%%%%%%%%%%%%%%%%%%%%%%%%%%%%
\section{Experimental Set-up and Observables}
\label{sec:setup}
%%%%%%%%%%%%%%%%%%%%%%%%%%%%%%%%%%%%%%%%%%%%%%%%%%%%%%%%%%%%%%%%%%%%%%

Many authors~\cite{nufactory} have emphasized  the advantages of using 
the long straight section of a high intensity muon storage ring to make a 
neutrino factory. 
The muons (anti-muons) accelerated to an energy $E_\mu$
constitute a pure source of both $\nu_\mu$ ($\bar \nu_\mu$) and $\bar \nu_e$ 
($ \nu_e$) through their decay $\mu^- \to e^- \bar \nu_e \nu_\mu$ 
($\mu^+ \to e^+  \nu_e \bar \nu_\mu$) with well known initial flux and 
energy distribution. 
In this context it is quite suitable to perform 
very precise measurements of 
the probability of oscillation in $CP$-conjugated channels, such as 
$\nu_\mu \to \nu_\tau$   and $\bar \nu_\mu \to \bar \nu_\tau$ or 
$\nu_e \to \nu_\tau$   and $\bar \nu_e \to \bar \nu_\tau$.

There are many propositions for these type of Neutrino Factories, with 
values for the energy of the stored muon (anti-muon),  $E_\mu$, going  from 
10 GeV  to 250 GeV, for baselines, $L$, ranging from 730 km to 10\, 000 km,  
in any case the neutrino beam penetrating  a fair bit of the Earth's crust.
Since we do not know which will be the final configuration we will 
do our estimations for several possible configurations. 
We do our calculations taking into account all the available relevant 
experimental information to compute real experimental observables. 

Here we will explore the neutrino factory as an appearance 
$\nu_\tau$ ($\bar \nu_\tau$) experiment.  
We define the following observables of interest:

\begin{equation}
R_{\mu\tau} (\epsilon_{\mu \tau},\epsilon^\prime_{\mu \tau})= 
\frac{N^+_{\bar \nu_\tau}(\epsilon_{\mu \tau},\epsilon^\prime_{\mu \tau})}{N_{\nu_\tau}^-(\epsilon_{\mu \tau},\epsilon^\prime_{\mu \tau})},
\label{rat1}
\end{equation}
and 
\begin{equation}
R_{e\tau} (\epsilon_{e \tau}, \epsilon^\prime_{e \tau})= \frac{N^-_{\bar \nu_\tau}(\epsilon_{e \tau},\epsilon^\prime_{e \tau})}{N^+_{\nu_\tau}(\epsilon_{e \tau},\epsilon^\prime_{e \tau})},
\label{rat2}
\end{equation}
where $R_{\mu \tau}$ ($R_{e \tau}$) is the ratio between 
the total number of detectable $\bar \nu_\tau$-events, 
$N^+_{\bar \nu_\tau}$ ($N^-_{\bar \nu_\tau}$), 
when the neutrino beam is made of $\mu^+$ ($\mu^-$) decays,
in the $\bar{\nu}_\mu \to \bar{\nu}_\tau$  
($\bar{\nu}_e \to \bar{\nu}_\tau$) channel, 
over the total number of detectable $\nu_\tau$-events, 
$N^-_{\nu_\tau}$ ($N^+_{\nu_\tau}$),
when the beam is made of  $\mu^-$ ($\mu^+$) decays, 
in the $\nu_\mu \to \nu_\tau$  ($\nu_e \to \nu_\tau$) channel. 
These numbers can be calculated as

\begin{equation}
N^-_{\nu_\tau}(\epsilon_{\mu \tau},\epsilon^\prime_{\mu \tau}) = n_{\mu^-} M \frac{10^9 N_{\text{A}}}{m^2_\mu \pi} \frac{E^3_\mu}{L^2} \int_{E_{\text{th}}}^{E_{\mu}} 
h(E) P(\nu_\mu \to \nu_\tau) (E)  dE,
\label{mutau}
\end{equation}

\begin{equation}
N^+_{\bar \nu_\tau}(\epsilon_{\mu \tau},\epsilon^\prime_{\mu \tau}) = n_{\mu^+} M \frac{10^9 N_{\text{A}}}{m^2_\mu \pi} \frac{E^3_\mu}{L^2} \int_{E_{\text{th}}}^{E_{\mu}} 
\bar h(E) P(\bar \nu_\mu \to \bar\nu_\tau) (E)  dE,
\label{anti-mutau}
\end{equation}

\begin{equation}
N^+_{\nu_\tau}(\epsilon_{e \tau},\epsilon^\prime_{e \tau}) = n_{\mu^-} M \frac{10^9 N_{\text{A}}}{m^2_\mu \pi} \frac{E^3_\mu}{L^2} \int_{E_{\text{th}}}^{E_{\mu}} 
g(E) P(\nu_e \to \nu_\tau) (E)  dE,
\label{etau}
\end{equation}

\begin{equation}
N^-_{\bar \nu_\tau}(\epsilon_{e \tau},\epsilon^\prime_{e \tau}) = n_{\mu^+} M \frac{10^9 N_{\text{A}}}{m^2_\mu \pi} \frac{E^3_\mu}{L^2} \int_{E_{\text{th}}}^{E_{\mu}} 
\bar g(E) P(\bar \nu_e \to \bar\nu_\tau) (E)  dE,
\label{anti-etau}
\end{equation}

with 
\begin{eqnarray}
h(E) = & \displaystyle 2\frac{E^2}{E_\mu^2}(3-2\frac{E}{E_\mu}) \frac{\sigma_{\nu_{\tau}}(E)}{E_\mu^2} \eta_{\tau}(E), \\
\bar h(E) = &  \displaystyle 2\frac{E^2}{E_\mu^2}(3-2\frac{E}{E_\mu}) \frac{\sigma_{\nu_{\bar \tau}}(E)}{E_\mu^2}  \,\eta_{\bar \tau}(E), \\
g(E) = & \displaystyle 12\frac{E^2}{E_\mu^2}(1-\frac{E}{E_\mu}) \frac{\sigma_{\nu_{\tau}}(E)}{E_\mu^2} \eta_{\tau}(E), \\
\bar g(E) = &  \displaystyle 12\frac{E^2}{E_\mu^2}(1-\frac{E}{E_\mu}) \frac{\sigma_{\nu_{\bar \tau}}(E)}{E_\mu^2}  \,\eta_{\bar \tau}(E), \\
\end{eqnarray}
where $E_\mu$ is the muon source energy, $M$ is the detector mass 
in ktons,  
$n_{\mu^-}$ and $n_{\mu^+}$ the number of useful 
$\mu^-$ and $\mu^+$ decays, 
respectively, $10^{9} N_{\text{A}}$ is the number of 
nucleons in a kton and $m_\mu$ is the mass of the muon.  
The functions $h(E)$ and $\bar h(E)$ ($g(E)$ and $\bar g(E)$), contain 
the $\nu_\mu$ and $\bar \nu_\mu$ ($\nu_e$ and $\bar \nu_e$)
energy spectrum normalized to 1, the charged current interaction cross 
section for  $\nu_\tau$ and $\bar \nu_\tau$~\cite{x-section},  and  
the $\nu_\tau$, $\bar \nu_\tau$  detection efficiencies 
$\eta_{\tau}$,$\eta_{\bar \tau}$, respectively.
 
We have used $E_{\text{th}} = 4$ GeV and assumed, for simplicity, that once 
this cut is applied the tau neutrino and tau anti-neutrino, can
be observed with the same efficiency 
$\eta_{\tau}=\eta_{\bar \tau}=0.33$ which seems to be 
achievable by means of exploiting one-prong and three-prong $\tau$ decay 
topologies along with displaced vertex or kinks resulting from 
$\tau$-lepton decays~\cite{fnf}.
We have neglected the finite detector resolution following Ref.\ \cite{flpr}. 

In our study we will consider the following general characteristics for the 
neutrino factory~: $E_\mu=20$ GeV and 50 GeV, $2 \times 10^{20}$ muon decays, 
a 10 kton detector and 5 years of data taking. We would like 
to point out that the observable  
$R_{\mu \tau}$ ($R_{e \tau}$) can amplify the differences 
between $N^-_{\nu_\tau}(N^+_{\nu_\tau})$ and 
$N^+_{\bar{\nu}_\tau}(N^-_{\bar{\nu}_\tau})$, independent 
of the absolute number of events.

%%%%%%%%%%%%%%%%%%%%%%%%%%%%%%%%%%%%%%%%%%%%%%%%%%%%%%%%%%%%%%%%%%%%%%
\section{Perspective of Future Neutrino Factories}
\label{sec:res}
%%%%%%%%%%%%%%%%%%%%%%%%%%%%%%%%%%%%%%%%%%%%%%%%%%%%%%%%%%%%%%%%%%%%%%

In order to estimate the maximal limit on $\epsilon_{\alpha \tau}$ and 
$\epsilon^\prime_{\alpha \tau}$, with $\alpha=e,\mu$ that can be achieved 
by a neutrino factory we define the following functions 

\bea
\chi^2(\epsilon_{\alpha \tau},\epsilon^\prime_{\alpha \tau}) &=& 
2 [N^{\pm}_{\nu_\tau}(\epsilon_{\alpha \tau},\epsilon^\prime_{\alpha \tau})-
N^{\pm}_{\nu_\tau}(\epsilon_{\alpha \tau}=0,\epsilon^\prime_{\alpha \tau}=0)] \nonumber \\
&+& 2 N^{\pm}_{\nu_\tau}(\epsilon_{\alpha \tau}=0,\epsilon^\prime_{\alpha \tau}=0) \; \ln \left( \frac{N^{\pm}_{\nu_\tau}(\epsilon_{\alpha \tau}=0,\epsilon^\prime_{\alpha \tau}=0)}
{N^{\pm}_{\nu_\tau}(\epsilon_{\alpha \tau}, \epsilon^\prime_{\alpha \tau})}\right),
\label{conf-l1}
\eea
and
\bea
\bar \chi^2(\epsilon_{\alpha \tau},\epsilon^\prime_{\alpha \tau}) &=& 
2 [N^{\mp}_{\bar \nu_\tau}(\epsilon_{\alpha \tau},\epsilon^\prime_{\alpha \tau})-
N^{\mp}_{\nu_\tau}(\epsilon_{\alpha \tau}=0,\epsilon^\prime_{\alpha \tau}=0)] \nonumber \\
&+& 2 N^{\mp}_{\bar \nu_\tau}(\epsilon_{\alpha \tau}=0,\epsilon^\prime_{\alpha \tau}=0) \; \ln \left( \frac{N^{\mp}_{\bar \nu_\tau}(\epsilon_{\alpha \tau}=0,\epsilon^\prime_{\alpha \tau}=0)}
{N^{\mp}_{\bar \nu_\tau}(\epsilon_{\alpha \tau}, \epsilon^\prime_{\alpha \tau})}\right),
\label{conf-l2}
\eea
so that for fixed values of the oscillation parameters, one can compute the 
number of standard deviations of separation between pure mass induced 
oscillation and mass induced oscillation plus NSNI contribution as a function 
of $\epsilon_{\alpha \tau}$ and $\epsilon^\prime_{\alpha \tau}$, as 
$n_\sigma = \sqrt{(\chi^2+\bar \chi^2)/2}$.    

We will consider in our analyses the following range for the oscillation 
parameters : 
$1 \times 10^{-3}$ eV$^2$ $\lsim $ 
$|\Delta m^2_{31}| \lsim 7 \times 10^{-3}$ eV$^2$,  
$|\Delta m^2_{21}| \lsim 1 \times 10^{-3}$ eV$^2$, 
$0.8 \lsim \sin^2 2 \theta_{23} \lsim  1.0$,   
$0.7 \lsim \sin^2 2 \theta_{12} \lsim 1.0$ and 
$\sin^2 2 \theta_{13} \lsim 0.3$; with 
$0 \leq \theta_{12}, \theta_{23}, \theta_{13} < \pi/2$ and 
$0 \leq \delta \leq \pi/2$. 
To the extent that the contributions from the sub-leading solar 
$|\Delta m^2_{21}| << |\Delta m^2_{31}|$ scale are small, our results apply 
approximately to other solar scenarios.

\subsection{NSNI in $\nu_\mu \to \nu_\tau$ channel}

The interesting feature of this channel is not only that for the 
standard oscillation mechanism, 
with $\delta=0$, the fake $CP$ violation 
induced by matter effect is very small~\cite{CPV_matter}, 
under our assumptions about 
the mixing parameters, but also that 
even if we consider the $CP$ violation phase 
to be maximal ($\delta = \pi/2$), this quantity remains quite 
negligible. This is due to the fact that $\sin^2 2 \theta_{13}$ is highly
constrained by CHOOZ and the atmospheric neutrino data. On the 
other hand,  as we have seen in Sec.\ \ref{sec:form} the inclusion of 
an extra contributions from NSNI in the neutrino evolution Hamiltonian 
will enhance the fake $CP$ violation in matter.  

\begin{figure}
\vglue -1.0cm
\hglue 1cm
\centerline{
\epsfig{file=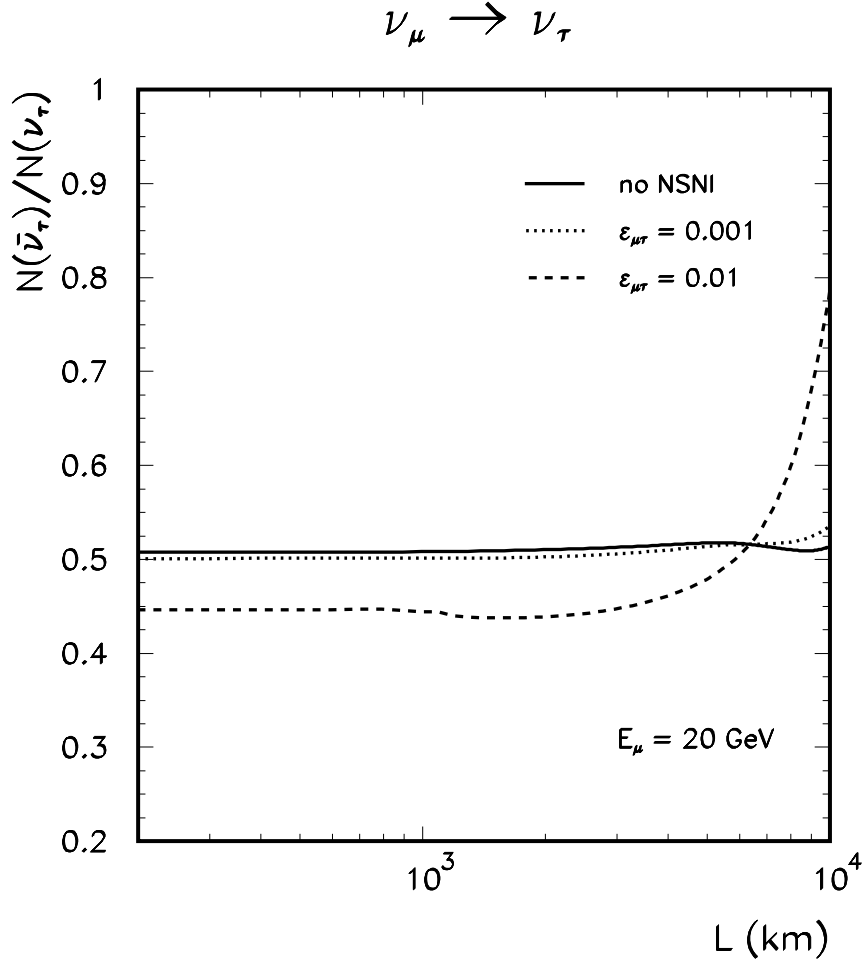,height=20.0cm,width=20.cm}}
\vglue -4.5cm
\caption{Ratio between the number of expected $\bar \nu_\tau$ 
over the number of expected $\nu_\tau$, as a function of the baseline $L$, 
for a few values of $\epsilon_{\mu \tau}$ and 
$\epsilon_{\mu \tau}^\prime=0$. 
The value of the other oscillation 
parameters used  in this plot are : $\Delta m^2_{32} = 3.1 \times 10^{-3}$
eV$^2$, $\Delta m^2_{21} = 3.3 \times 10^{-5}$ eV$^2$, 
$\sin^2 2\theta_{23}= 0.97$, $\sin^2 2\theta_{12}= 0.78$,
$\sin^2 2\theta_{13}= 0.02$ and $\delta=0$. 
}
\label{fig1}
\vglue -1.cm
\end{figure}

\begin{figure}
\vglue -1.0cm
\hglue 1cm
\centerline{
\epsfig{file=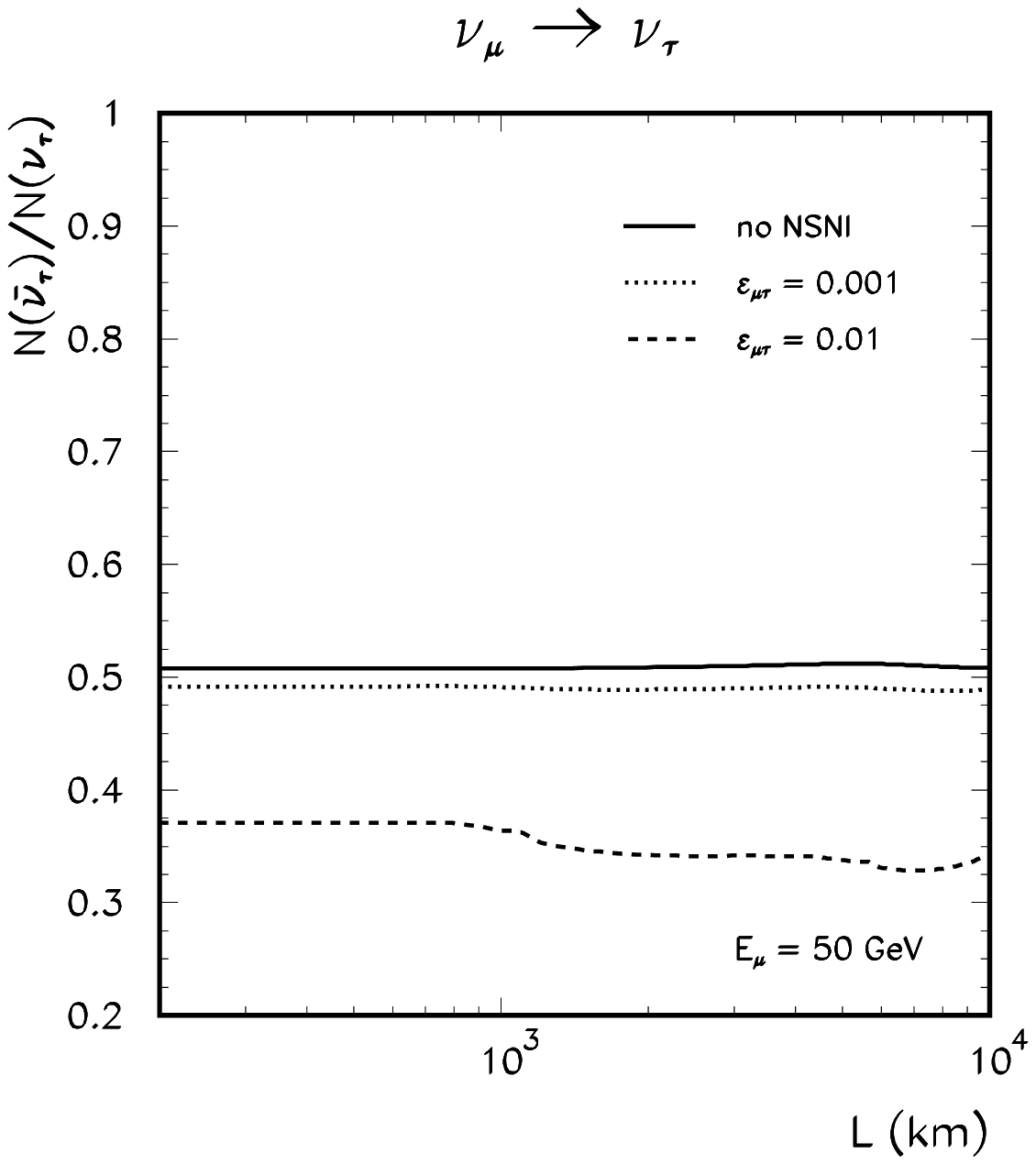,height=20.0cm,width=20.cm}}
\vglue -4.5cm
\caption{Same as Fig.\ \ref{fig1} but for $E_\mu$ = 50 GeV. }
\label{fig2}
\vglue -1.cm
\end{figure}

To illustrate the size of the effect due to NSNI in this channel we plot in 
Fig. \ref{fig1} and \ref{fig2} the  ratio $R_{\mu \tau}$ as a function of $L$
for $E_\mu=20$ and 50 GeV respectively, for 
$\epsilon^\prime_{\mu \tau}=0$ and $\epsilon_{\mu \tau} = 0.001$ and 0.01.
These plots were done for the best fit values of the oscillation parameters 
for the combined analysis of atmospheric, solar and reactor data according 
to Ref.~\cite{cmpv}, assuming the large mixing angle solution to the solar 
neutrino anomaly, 
\ie, $\Delta m^2_{31} = 3.1 \times 10^{-3}$ eV$^2$,  
$\Delta m^2_{21} = 3.3 \times 10^{-5}$ eV$^2$, 
$\sin^2 2 \theta_{23} =  0.97$,   
$\sin^2 2 \theta_{12} = 0.78$ and
$\sin^2 2 \theta_{13} = 0.02$, with $\delta=0$. 

For no NSNI, the ratio is nearly constant around 0.5, it is essentially 
dominated by the ratio between $\nu_\tau$ and $\bar \nu_\tau$ cross sections
since the standard matter effect is negligible 
($P(\nu_\mu \to \nu_\tau) \approx P(\bar \nu_\mu \to \bar \nu_\tau)$) .
As FCNI are introduced by increasing the value of $\epsilon_{\mu \tau}$, 
we observe the effect of the new interactions pop in, becoming quite 
appreciable for $\epsilon_{\mu \tau} \sim 10^{-2}$ and stronger as we 
increase in energy. This dependence in energy can be easily understood 
using the two generation formula given in 
Eqs.~(\ref{prob2g})-(\ref{ratio_approx}), 
setting $\Delta= \Delta_{32}$ and $\sin 2\theta=\sin 2 \theta_{23}$. 
Although we are here working in a full three generation scheme, this 
approach will be good enough to interpret the general behavior of the 
probabilities. This is because these two parameters are the leading 
parameters for conversion in this channel in the three generation framework.
>From Eqs. (\ref{prob2g_approx}) and (\ref{ratio_approx}), 
we see that the effect of fake $CP$ violation become larger 
as energy grows and can reach smaller values of 
$\epsilon_{\mu \tau}$.

Note that the impact of using the Earth's density profile can be appreciated 
in the small distortions of the curves for $L> 1000$ km, which is specially 
sizable in Fig.~\ref{fig2}, for $\epsilon_{\mu \tau}$ =0.01, which also can 
be explained in first approximation by the two generation formula.

For our statistical analyses of the limits that can be achieved by a 
neutrino factory, we have chosen to work with $L=$ 732 km. Two reasons support 
this choice : the FCNI effect is stronger and independent of $L$ up to 1000 
km (which can be understood from Eq. (\ref{ratio_approx}) 
in Sec.~\ref{useful})
and this baseline is compatible with the CERN-Gran Sasso 
or Fermilab-Soudan distance.
Besides this at 732 km the neutrino flux is still quite big. 

\begin{figure}
\vglue -1.0cm
\hglue 1cm
\centerline{
\epsfig{file=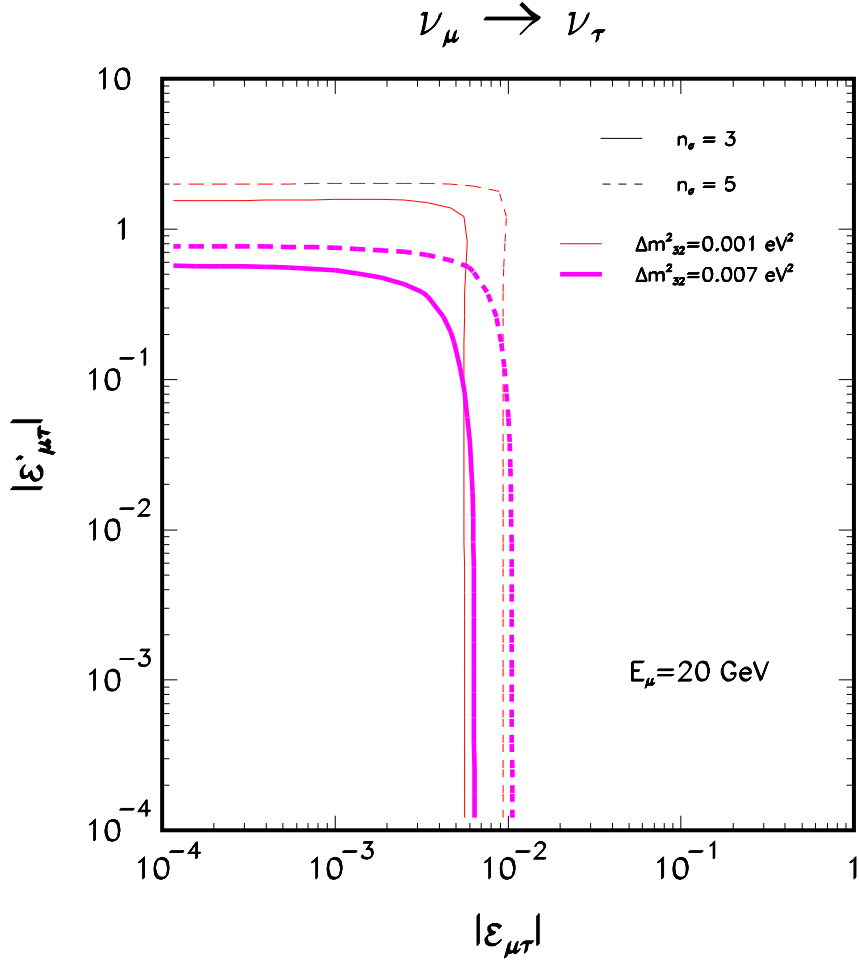,height=20.0cm,width=20.cm}}
\vglue -4.5cm
\caption{Region in the $\epsilon^\prime_{\mu \tau} \times \epsilon_{\mu \tau}$
plane that can be probed by a neutrino factory considering $E_\mu=20$ GeV, 
$L=732$ km, $2 \times 10^{20}$ muon decays, a 10 kton detector after 5 years 
of data taking. The regions that were determined by 
demanding a $n_\sigma =3,5$ separation, are the ones to the right of 
each line.  The value of the other oscillation parameters used  in this plot 
are : $\Delta m^2_{21} = 3.3 \times 10^{-5}$ eV$^2$, 
$\sin^2 2\theta_{23}= 0.97$, $\sin^2 2\theta_{12}= 0.78$,
$\sin^2 2\theta_{13}= 0.02$ and $\delta=0$. 
Here $\epsilon_{\mu \tau},\epsilon_{\mu \tau}^\prime>0$. 
}
\label{fig3}
\vglue -1.cm
\end{figure}

\begin{figure}
\vglue -1.0cm
\hglue 1cm
\centerline{
\epsfig{file=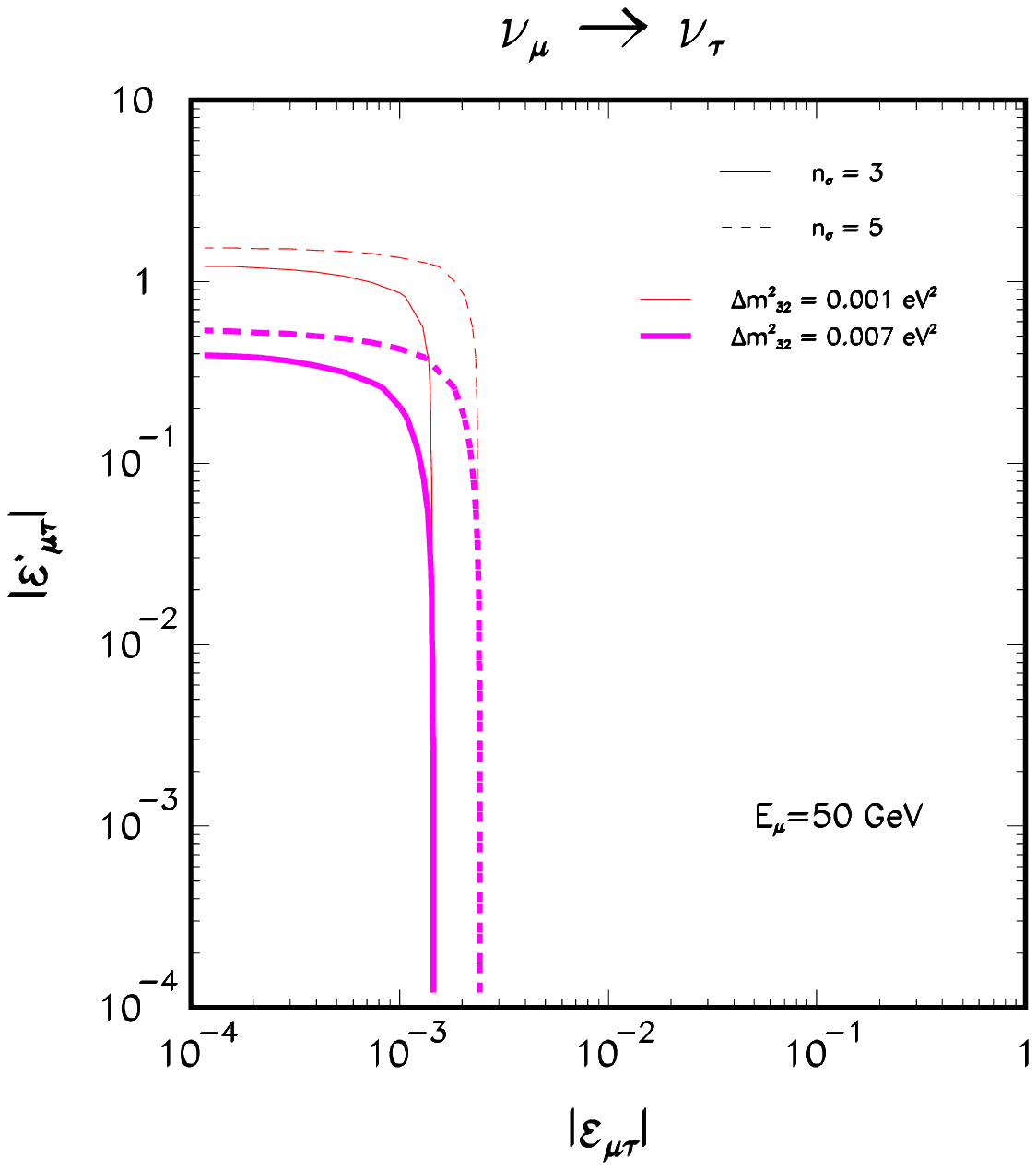,height=20.0cm,width=20.cm}}
\vglue -4.5cm
\caption{Same as Fig.\ \ref{fig3} but for $E_\mu=50$ GeV.}
\label{fig4}
\vglue -1.cm
\end{figure}

In Figs.~\ref{fig3} and \ref{fig4}, we show the region of
sensitivity in 
$|\epsilon^\prime_{\mu \tau}| \times |\epsilon_{\mu \tau}|$ plane 
for $\epsilon^\prime_{\mu \tau},\epsilon_{\mu \tau}>0$ computed 
using the prescription given in Sec.~\ref{sec:res} and requiring a 
$n_\sigma=3,5$ separation. We plot the maximal limit for two extreme cases, 
one for $\Delta m^2_{32}=1 \times 10^{-3}$ eV$^2$ and the other for
$\Delta m^2_{32}=7 \times 10^{-3}$ eV$^2$.
To give an idea of the corresponding number of events  expected at the 
limiting points, we give in Tab.~\ref{tab1} these numbers
for $\epsilon^\prime_{\mu \tau}=0$. 
We have checked that these results are not modified by changes in the  
values of the other oscillation parameters or by the adopted sign 
of $\Delta m^2_{32}$ or of $\Delta m^2_{21}$.
In particular, $P(\nu_\mu \to \nu_\tau)$ and the 
$P(\bar \nu_\mu \to \bar \nu_\tau)$ are essentially independent of $\delta$ 
and the value of $\Delta m^2_{21}$. 
We also have done the same calculation in the case of either  
$\epsilon^\prime_{\mu \tau}<0$ or $\epsilon_{\mu \tau}<0$ or both, 
but we do not show these curves here since they are almost 
identical to the ones in Figs.~\ref{fig3} and \ref{fig4}.   

For $E_\mu=$ 20 GeV we see that it is possible to probe, at 3 $\sigma$ level,
$|\epsilon_{\mu \tau}| \gsim 6 \times 10^{-3}$  and 
$|\epsilon^\prime_{\mu \tau}| \gsim 1.7~(0.6)$,  
for $E_\mu=$ 50 GeV, $|\epsilon_{\mu \tau}| \gsim 1.6 \times 10^{-3}$ 
and $|\epsilon^\prime_{\mu \tau}| \gsim 1.4~(0.4)$, for $\Delta m^2_{32} = 1 \times 10^{-3}$ eV$^2$ ($7 \times 10^{-3}$ eV$^2$).  
As explained in Sec.~\ref{sec:res} we did not expect to be very 
sensitive to $\epsilon^\prime_{\mu \tau}$, also the  limits are better at 
50 GeV due to the flux increase with $E_\mu^3$.

Finally we note that, although the oscillation probabilities in this channel 
 can get quite large, the difference in the neutrino and anti-neutrino 
probabilities remains small, unless $|\epsilon_{\mu \tau}| \to 1$. 
This means that the constraints relies mainly on a large event statistics.

\subsection{NSNI in $\nu_e \to \nu_\tau$ channel}

\begin{figure}
\vglue -1.0cm
\hglue 1cm
\centerline{
\epsfig{file=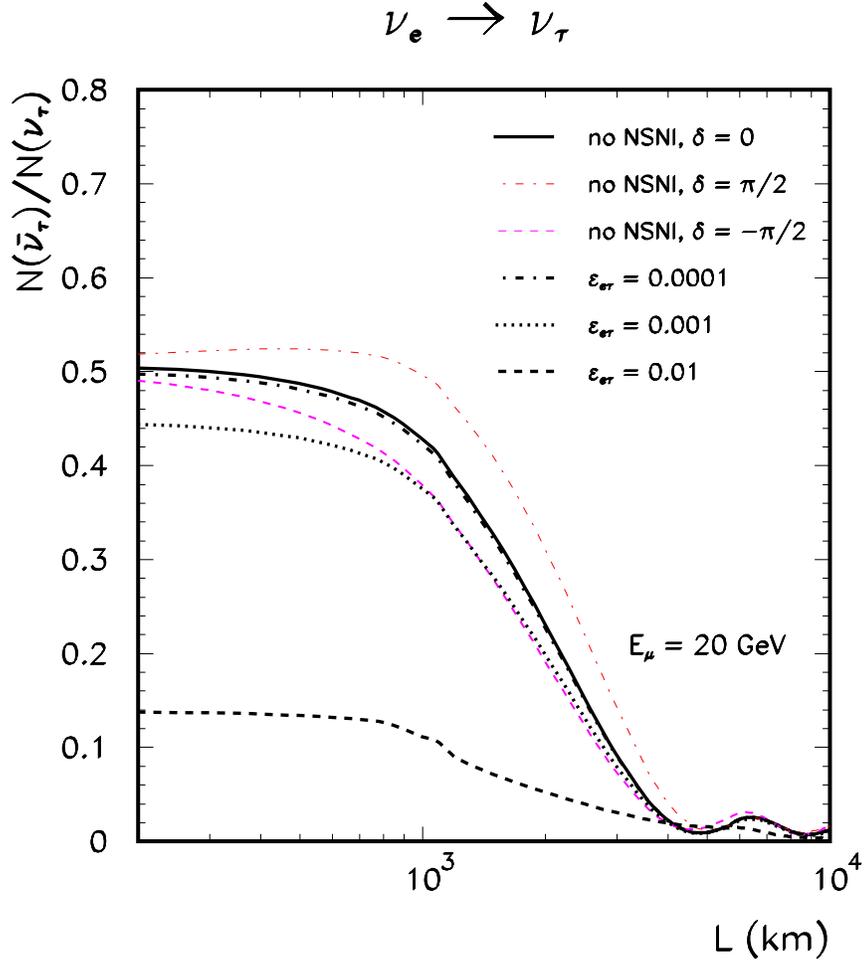,height=20.0cm,width=20.cm}}
\vglue -4.5cm
\caption{Same as Fig.\ \ref{fig1} but for the mode 
$\nu_e \to \nu_\tau$/$\bar \nu_e \to \bar \nu_\tau$, we also 
show the standard case for $\delta=\pi/2,-\pi/2$.}
\label{fig5}
\vglue -1.cm
\end{figure}

\begin{figure}
\vglue -1.0cm
\hglue 1cm
\centerline{
\epsfig{file=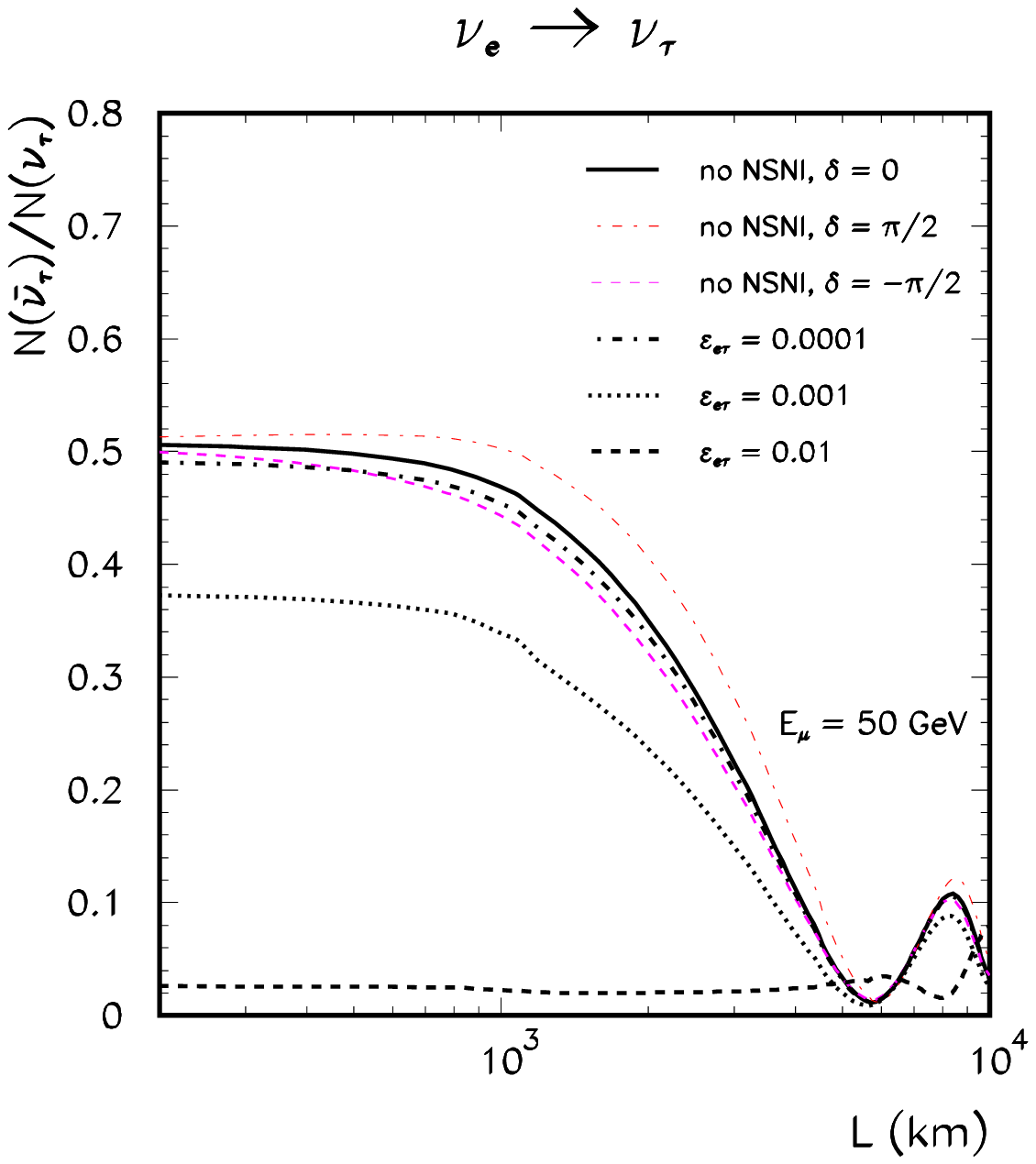,height=20.0cm,width=20.cm}}
\vglue -4.5cm
\caption{Same as Fig.\ \ref{fig5} but for $E_\mu$ = 50 GeV. }
\label{fig6}
\vglue -1.cm
\end{figure}

We plot in Figs.~\ref{fig5} and  \ref{fig6} the ratio $R_{e \tau}$ as a 
function of $L$ for $E_\mu=$ 20 GeV and 50 GeV, respectively, for 
$\epsilon^\prime_{e \tau}=0$ and $\epsilon_{e \tau} = 0.0001, 0.001$ and 0.01.
We can compare these figures with Figs.~\ref{fig1} and \ref{fig2} 
for $R_{\mu \tau}$ and make two observation. 
First, for the case without NSNI we have a strong 
deviation in the channel $\nu_e \to \nu_\tau$
from the value 0.5 after $L \gsim 1000 $ km, 
this is because matter effects ($V_e$) are specially important in this mode, 
in contrast with $\nu_\mu \to \nu_\tau$ where they 
are small. Second, when we add the non-standard contributions the effect 
in $\nu_e \to \nu_\tau$ channel is again more relevant than for 
$\nu_\mu \to \nu_\tau$. This can be understood since at first approximation 
we can think that the modified three generation probabilities 
$P(\nu_e \to \nu_\tau)$ and $P(\bar \nu_e \to \bar \nu_\tau)$, 
as it is the case for the standard ones~\cite{bgrw},  must be proportional 
to $\sin^2 2 \theta_{13}^m$ in matter, only in our case we also have to 
consider the contributions of the non-standard interactions.
This permits a qualitative understanding of our results in terms of 
the two-generation formula (Eq.~(\ref{prob2g})) replacing $\Delta$ by 
$\Delta_{31}$ and $\sin 2 \theta$ by $\sin 2 \theta_{13}$, which are the 
most relevant parameters in this mode.
>From Eqs.~(\ref{prob2g_approx}) and (\ref{ratio_approx}), 
we see that the relative effect of fake $CP$ violation 
become larger compared to the $\nu_\mu \to  \nu_\tau$ 
channel because of the smaller value of the relevant 
mixing angle $\theta_{13}$ for this channel.

\begin{figure}
\vglue -1.0cm
\hglue 1cm
\centerline{
\epsfig{file=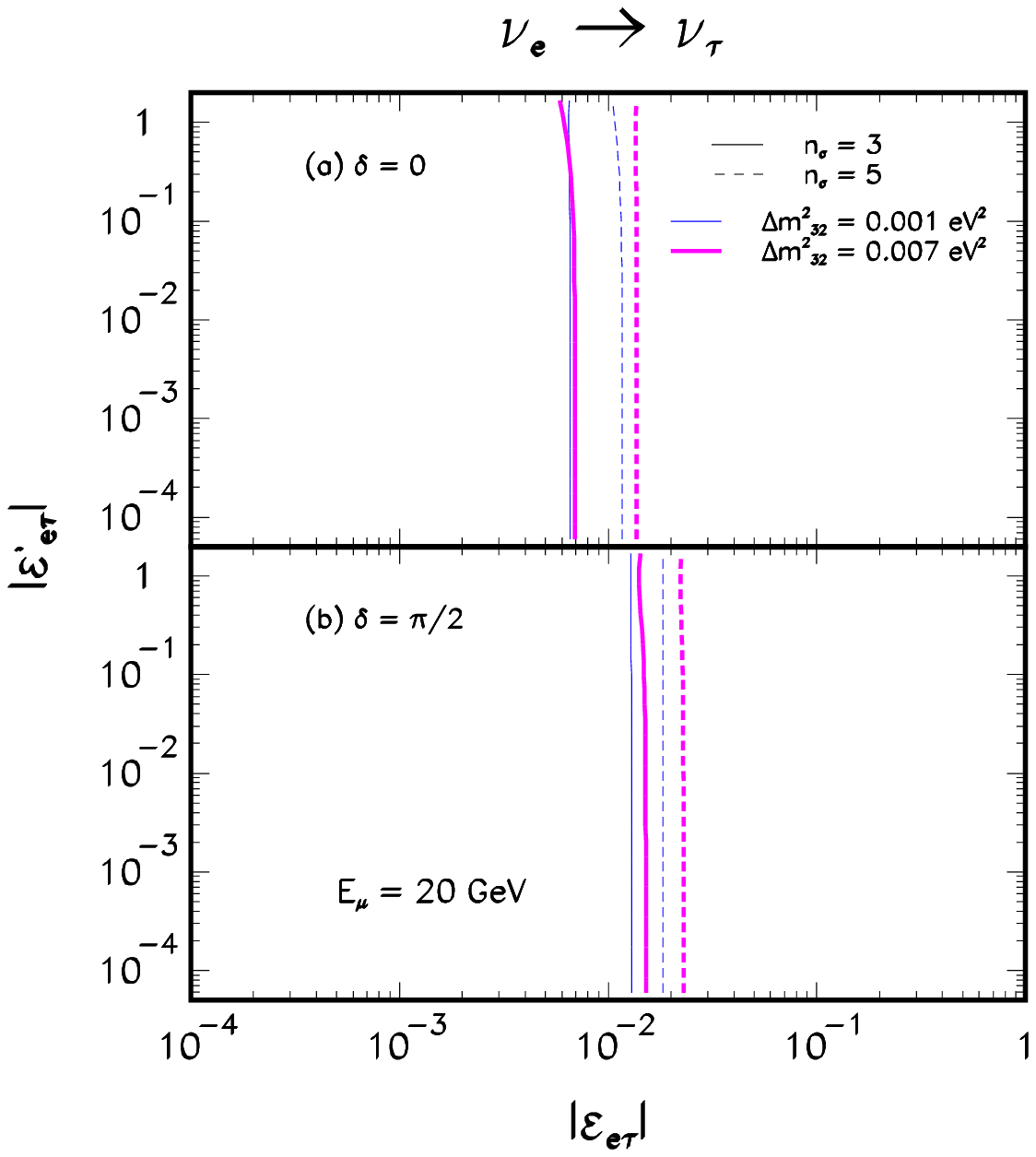,height=20.0cm,width=20.cm}}
\vglue -4.5cm
\caption{Same as Fig.\ \ref{fig3} but for the mode
$\nu_e \to \nu_\tau$/$\bar \nu_e \to \bar \nu_\tau$  and two distinct 
values of $\delta$. Here $\epsilon_{e \tau},\epsilon_{e \tau}^\prime>0$.}
\label{fig7}
\vglue -1.cm
\end{figure}

\begin{figure}
\vglue -1.0cm
\hglue 1cm
\centerline{
\epsfig{file=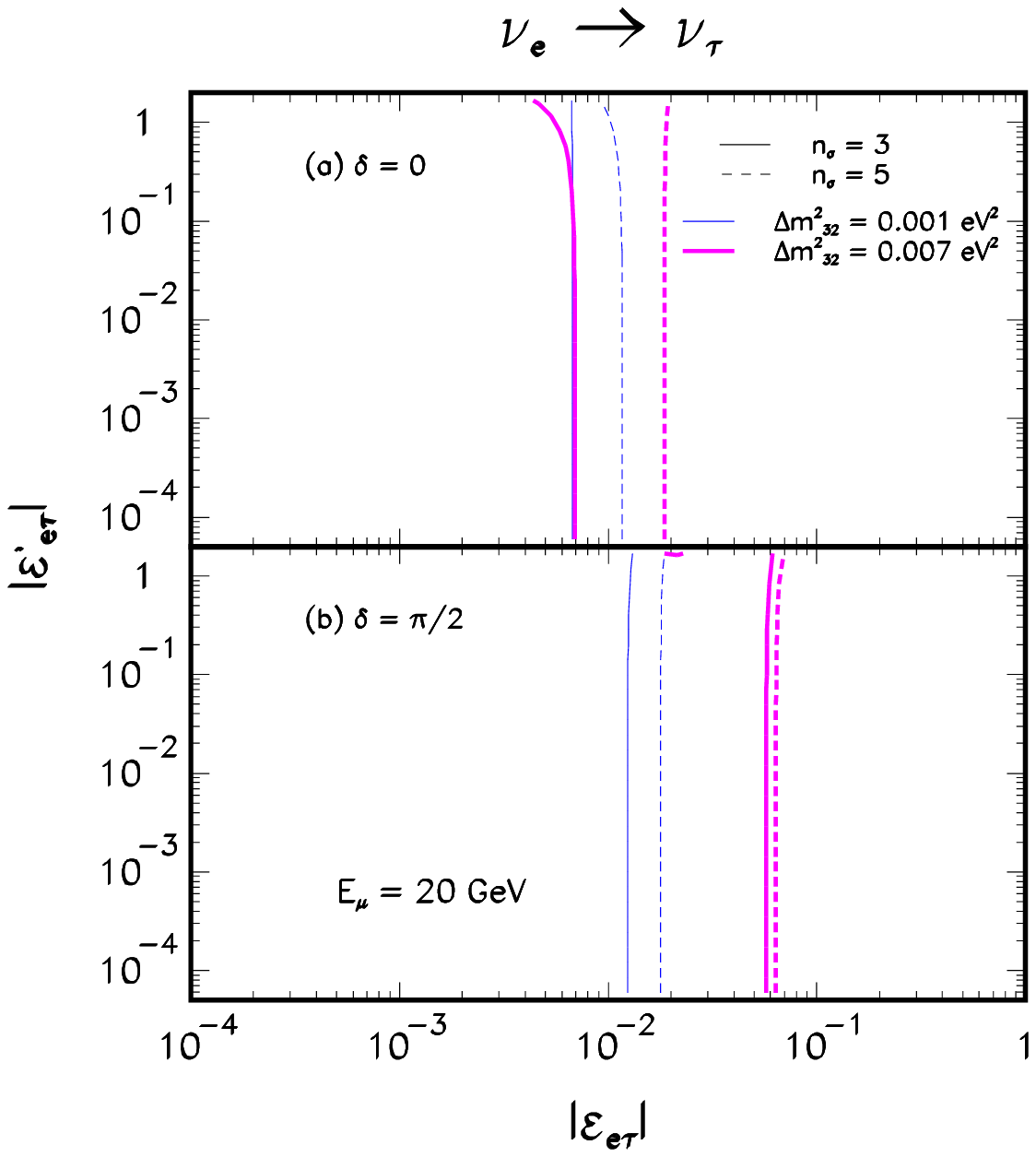,height=20.0cm,width=20.cm}}
\vglue -4.5cm
\caption{Same as Fig.\ \ref{fig3} but for the mode
$\nu_e \to \nu_\tau$/$\bar \nu_e \to \bar \nu_\tau$  and two distinct 
values of $\delta$. Here $\epsilon_{e \tau},\epsilon_{e \tau}^\prime<0$.}
\label{fig8}
\vglue -1.cm
\end{figure}

\begin{figure}
\vglue -1.0cm
\hglue 1cm
\centerline{
\epsfig{file=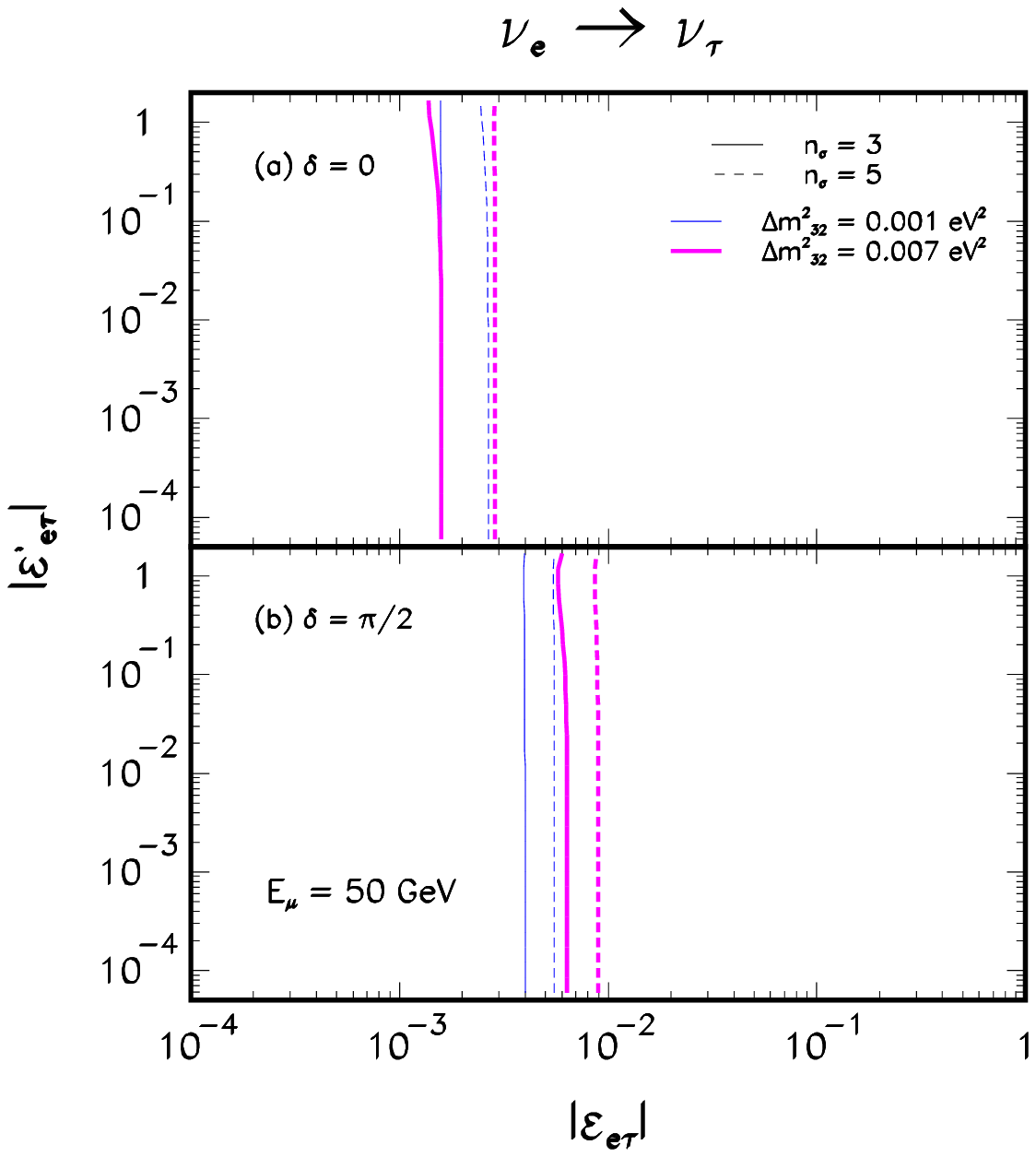,height=20.0cm,width=20.cm}}
\vglue -4.5cm
\caption{Same as Fig.\ \ref{fig7} but for $E_\mu=50$ GeV.}
\label{fig9}
\vglue -1.cm
\end{figure}

\begin{figure}
\vglue -1.0cm
\hglue 1cm
\centerline{
\epsfig{file=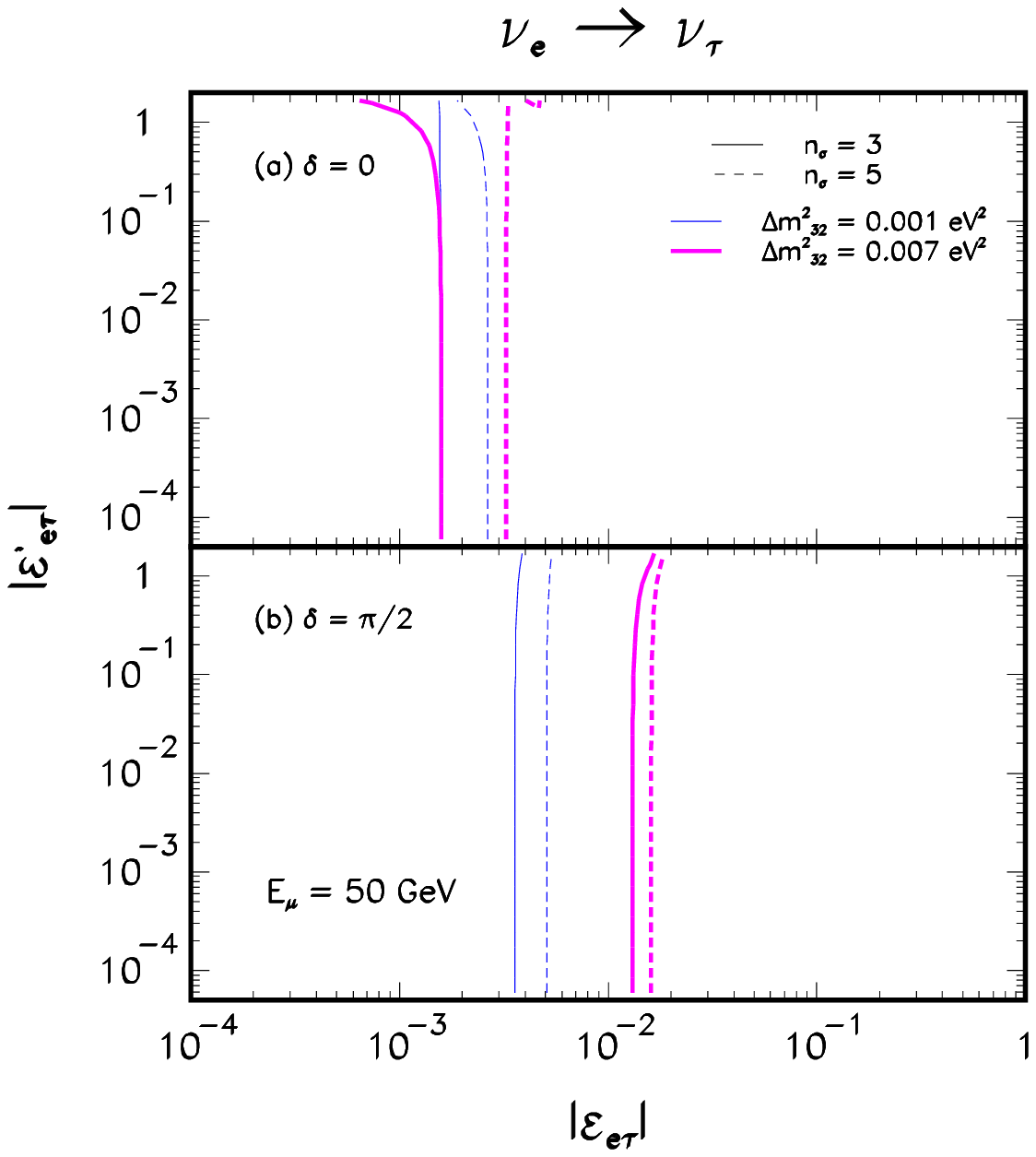,height=20.0cm,width=20.cm}}
\vglue -4.5cm
\caption{Same as Fig.\ \ref{fig8} but for $E_\mu=50$ GeV.}
\label{fig10}
\vglue -1.cm
\end{figure}

In Figs.~\ref{fig7}-\ref{fig10}, we show the minimal values 
of $\epsilon_{e\tau}$ that can be probed at  
$E_\mu=$  20 and 50 GeV.  The first two figures correspond to 
the cases $\epsilon_{e \tau},\epsilon^\prime_{e \tau}>0$ and 
$\epsilon_{e \tau},\epsilon^\prime_{e \tau}<0$ 
at 20 GeV, the last two to the cases 
$\epsilon_{e \tau},\epsilon^\prime_{e \tau}>0$ and 
$\epsilon_{e \tau},\epsilon^\prime_{e \tau}<0$  
at 50 GeV. We also have looked at the possibility of having 
 $\epsilon_{e \tau}>0$, $\epsilon^\prime_{e \tau}<0$ and  
$\epsilon_{e \tau}<0$, $\epsilon^\prime_{e \tau}>0$, the limits in the first 
case are almost identical to the ones shown in Figs.~\ref{fig7} and 
\ref{fig9}, while in the latter case very similar to Figs.~\ref{fig8} and 
\ref{fig10}.

These plots were calculated again assuming a comfortable separation of 
$n_\sigma=3,5$ in the $\chi^2$ analysis described in 
Sec.~\ref{sec:res} and $L = 732$ km as for the $\nu_\mu \to \nu_\tau$ mode.
For each plot we have two different assumptions (a) $\delta=0$ and 
(b) $\delta=\pi/2$, in both we have considered the 
uncertainty on $\Delta m^2_{32}$.  We see clearly the $CP$ violation phase 
$\delta$ is significant in this channel.
We do not show the dependence of these limits on the values of the other 
oscillation parameters or on the sign of the mass squared  differences, 
since these would practically unaffect the plots.
In the Tab.~\ref{tab1}, one can find  the corresponding number of events
when $\epsilon_{e \tau}>0$ and $\epsilon^\prime_{e \tau}=0$ for 
$n_\sigma =3$.

For $E_\mu=$ 20 GeV, we see that it is possible to probe, 
at the 3 $\sigma$ level,  for any value of $\epsilon^\prime_{e \tau}$, 
$\epsilon_{e \tau} \gsim 7 \times 10^{-3}$ and 
$\epsilon_{e \tau} \lsim - 7 \times 10^{-3}$ 
for $\delta=0$. For $\delta = \pi/2$  one can test 
$\epsilon_{e \tau} \gsim 1.2\times 10^{-2}$ (~$1.5 \times 10^{-2}$) 
and $\epsilon_{e \tau} \lsim -1.4 \times 10^{-2}$ (~$-5.1 \times 10^{-2}$ ), 
these limits correspond to $\Delta m^2_{32} = 1 \times 10^{-3}$~eV$^2$ 
($7 \times 10^{-3}$~eV$^2$).

For $E_\mu=$ 50 GeV, it is possible to test
$\epsilon_{e \tau} \gsim 1.6\times 10^{-3}$ and 
$\epsilon_{e \tau} \lsim -1.6\times 10^{-3}$  for $\delta=0$ 
and  $\epsilon_{e \tau} \gsim 4.0\times 10^{-3}$ (~$6.3 \times 10^{-3}$ ) 
and $\epsilon_{e \tau} \lsim -3.6\times 10^{-3}$ (~$-1.4 \times 10^{-2}$ ) 
for $\delta=\pi/2$, again these limits correspond to  
$\Delta m^2_{32} = 1 \times 10^{-3}$~eV$^2$ ($7 \times 10^{-3}$~eV$^2$).
In this channel it will be not possible to set a stiff limit on 
$\epsilon^\prime_{e \tau}$  due to low statistics.

In spite of the low statistic of events in this mode, the difference in the 
probabilities of oscillation for neutrinos and anti-neutrinos in matter, 
as one can see from Figs.~\ref{fig5} and \ref{fig6},
makes it possible to achieve a sensitivity to FCNI as low as the previous mode.

%%%%%%%%%%%%%%%%%%%%%%%%%%%%%%%%%%%%%%%%%%%%%%%%%%%%%%%%%%%%%%%%%%%%%%
\section{Final Conclusions}
\label{sec:conc}

We have studied the effect of non-standard neutrino matter 
interactions which are assumed to be sub-dominant in 
the standard mass induced neutrino oscillation in 
a future neutrino factory experiments. 
When non-standard matter interactions is added to the 
mass oscillation mechanism, the difference between 
neutrino and anti-neutrino probabilities can be enhanced, 
causing fake $CP$ violating effect not coming from 
the intrinsic $CP$ phase.  

Based on this interesting feature, we have obtained,  
in $\nu_e \to \nu_\tau$ as well as  $\nu_\mu \to \nu_\tau$ 
oscillation modes,  possible limits on NSNI which can be 
accessible by the future neutrino factories, under the frame 
work of three neutrino mixing scheme, assuming the mixing 
parameters compatibles with the solar and atmospheric 
solution as well as reactor neutrino oscillation experiments.

We have shown that these facilities, with 5 years of 
operation and a 10 kt detector, are able to test FCNI 
down to a level of ($10^{-3}-10^{-2}$) $G_F$ in either 
$\nu_\mu \to \nu_\tau$ and $\nu_e \to \nu_\tau$ 
modes whereas sensitivity to FDNI is much worse
only to a level of ${\cal{O}}(1)$ $G_F$
for $\nu_\mu \to \nu_\tau$.
These limits on FCNI will certainly put much more stringent 
limits on the strength of lepton flavor violating 
new interactions, parameterized here by 
$\epsilon$, than the present bound can be found in 
the literature.~\cite{bgp,fc-sol}.

In particular, we conclude that if $E_\mu=$ 20 GeV and 
$L=$ 732 km, 
neutrino factory oscillation experiments will be
able to set limits on 
the FCNI parameters such as   
$|\epsilon^{u,d}_{\mu \tau}| \lsim 0.006$,  
$|\epsilon^{e}_{\mu \tau}| \lsim 0.018$, 
$-0.006 \lsim \epsilon^{u,d}_{e \tau} \lsim 0.007$,
$-0.018 \lsim \epsilon^{e}_{e \tau} \lsim 0.021$, 
and if $E_\mu=$ 50 GeV these limits can be 
$|\epsilon^{u,d}_{\mu \tau}| \lsim 0.002$,
$|\epsilon^{e}_{\mu \tau}| \lsim 0.006$, 
$ -0.014 \lsim \epsilon^{u,d}_{e \tau} \lsim 0.007$,
$-0.042 \lsim \epsilon^{e}_{e \tau} \lsim 0.021$. 
These bounds could be regarded as 
robust and model independent  
and can be used to constraint new physics in 
the electroweak sector. 
Let us also remark that, for both $\nu_\mu\to\nu_\tau$ 
and $\nu_e\to\nu_\tau$ channels, 
our bounds are mainly affected by the $|\Delta m^2_{31}|$ 
and do not practically depend on $|\Delta m^2_{21}|$ 
which is relevant for the solutions to the solar neutrino 
problem, and there is some effect from 
the $CP$ violation phase, $\delta$, for the 
$\nu_e \to \nu_\tau$ channel. 

The constraints on $\epsilon^\prime_{\mu \tau}$
that could be obtained are $\epsilon^{\prime u,d}_{\mu \tau} \lsim 2$ and 
$\epsilon^{\prime e}_{\mu \tau} \lsim 0.7$. As we 
had expected these limits are much weaker 
than the corresponding ones 
in $\epsilon_{\mu \tau}$, due to the fact that 
the modified probability (mass+NSNI) is 
slightly dependent in 
$\epsilon^\prime_{\mu \tau}$ (see 
discussion on sec II.C) for the 
distance consider. Also these constraints are clearly 
much less stringent than the ones in Refs.~\cite{bgp,fc-sol}. 
Due to low statistics no limit on $\epsilon^\prime_{e \tau}$ can be 
established.

We have not present the analysis for $\epsilon_{ e \mu}$ since it 
will be very hard to lower the bounds founded in the 
literature  ${\cal{O}}(10^{-5})$ by a neutrino factory. 

As we were finishing this work we came across Ref.~\cite{hep-new} 
where the effect of extra real as well as fake CP violation due to new
physics was studied in the context of oscillation experiments 
at neutrino factories. We note that the work in 
Ref.~\cite{hep-new} is somehow complementary to our since there, 
the effect of new interactions was considered in the process of 
production and detection of neutrinos, while here we only consider 
their effect in the neutrino propagation.

%%%%%%%%%%%%%% Thanks
\acknowledgments 
This work was supported by Funda\c{c}\~ao de Amparo \`a Pesquisa
do Estado de S\~ao Paulo (FAPESP) and by Conselho Nacional de e 
Ci\^encia e Tecnologia (CNPq).

%%%%%%%%%%%%%%%%%%%%%%%%%%%%%%%%%%%%%%%%%%%%%%%%%%%%%%%%%%%%%%%%%%%%%%%
%%%%%%%%%%%%%% References 
%%%%%%%%%%%%%%%%%%%%%%%%%%%%%%%%%%%%%%%%%%%%%%%%%%%%%%%%%%%%%%%%%%%%%%%

%%%%%%%%%%%%%%  Table I  %%%%%%%%%%%%%%%%%%%%%%%%%%%%%%%%%%%%%%%%%%%%%%
%
\vglue -2cm 
\begin{table}
\caption{\small Number of $\nu_\tau$ and $\bar \nu_\tau$  events expected
for two different $E_\mu$ values $L=$ 732 km, $2 \times 10^{20}$ $\mu$ decays, 
assuming a 10 kton detector after 5 years of data taking.
We give this information for $\Delta m^2_{32}= 1 \times 10^{-3}$ eV$^2$ 
and $7 \times 10^{-3}$ eV$^2$, keeping the other oscillation 
parameters fixed to $\Delta m^2_{21}= 3.3 \times 10^{-5}$ eV$^2$, 
$\sin^2 2\theta_{23}=0.97$, $\sin^2 2\theta_{12}=0.78$, 
$\sin^2 2\theta_{13}=0.02$ and $\delta=0$. Here $\epsilon^\prime_{\mu \tau} =\epsilon^\prime_{e \tau}=0$ and  $\epsilon_{\mu \tau},\,\epsilon_{e \tau}>0$.}  
\begin{center}
\begin{tabular}{c|c|c|c|c}
\hline
\multicolumn{5}{|c|}{$\nu_\mu \to \nu_\tau$}\\
\hline
\hline
\hline
\multicolumn{5}{c}{$E_\mu$= 20 GeV}\\
\hline
\hline
& \multicolumn{2}{c|}{$\Delta m_{32}^2 = 1 \times 10^{-3}$ eV$^2$} &\multicolumn{2}{c}{$\Delta m^2_{32} = 7\times 10^{-3}$ eV$^2$}  \\
 & $N_{\nu_\tau}$  & $N_{\bar \nu_\tau}$ & $N_{\nu_\tau}$ & $N_{\bar \nu_\tau}$ \\
\hline
no NSNI & 954  & 484  & 39\,372 & 20\,009 \\
$\epsilon_{\mu \tau}$  @ 3 $\sigma$ & 1\,061  &  433  & 40\,073  
 & 19\,656  \\
\hline 
\hline 
\multicolumn{5}{c}{$E_\mu$= 50 GeV}\\
\hline
\hline
& \multicolumn{2}{c|}{$\Delta m_{32}^2 = 1 \times 10^{-3}$ eV$^2$} &\multicolumn{2}{c}{$\Delta m^2_{32} = 7\times 10^{-3}$ eV$^2$}  \\
 & $N_{\nu_\tau}$  & $N_{\bar \nu_\tau}$ & $N_{\nu_\tau}$ & $N_{\bar \nu_\tau}$ \\
\hline
no NSNI & 2\,470 & 1\,253 & 114\,521 &  58\,153 \\ 
$\epsilon_{\mu \tau}$ @ 3 $\sigma$  & 2\,665 & 1\,156 & 115\,804 & 57\,506\\
\hline
\hline
\hline
\multicolumn{5}{|c|}{$\nu_e \to \nu_\tau$}\\
\hline
\hline
\hline
\multicolumn{5}{c}{$E_\mu$= 20 GeV}\\
\hline
\hline
& \multicolumn{2}{c|}{$\Delta m_{32}^2 = 1 \times 10^{-3}$ eV$^2$} &\multicolumn{2}{c}{$\Delta m^2_{32} = 7\times 10^{-3}$ eV$^2$}  \\
 & $N_{\nu_\tau}$ & $N_{\bar \nu_\tau}$ & $N_{\nu_\tau}$ & $N_{\bar \nu_\tau}$ \\
\hline
no NSNI & 6 & 3  &  384 & 163  \\
$\epsilon_{e \tau}$ @ 3 $\sigma$ & 18 & 1  & 456  &  134  \\
\hline
\hline
\multicolumn{5}{c}{$E_\mu$= 50 GeV}\\
\hline
\hline
& \multicolumn{2}{c|}{$\Delta m_{32}^2 = 1 \times 10^{-3}$ eV$^2$} &\multicolumn{2}{c}{$\Delta m^2_{32} = 7\times 10^{-3}$ eV$^2$}  \\
 & $N_{\nu_\tau}$ & $N_{\bar \nu_\tau}$ & $N_{\nu_\tau}$ & $N_{\bar \nu_\tau}$ \\
\hline
no NSNI & 15 & 8  &  1\,116 & 522  \\
$\epsilon_{e \tau}$ @ 3 $\sigma$ & 32 & 3  & 1\,235  &  468  \\
\hline
\end{tabular}
\label{tab1}
\end{center}
\end{table}

%%%%%%%%%%%%%%%%%%%%%%%%%%%%%%%%%%%%%%%%%%%%%%%%%%%%%%%%%%%%%%%%%%%%%%%
%%%%%%%%%%%%%% Begining of Figures
%%%%%%%%%%%%%%%%%%%%%%%%%%%%%%%%%%%%%%%%%%%%%%%%%%%%%%%%%%%%%%%%%%%%%%%

%%%%%%%%%%%%%%%%%%%%%%%%%%%%%%%%%%%%%%%%%%%%%%%%%%%%%%%%%%%%%%%%%%%%
%%%%%%%               Figure 1                           %%%%%%%%%%%
%%%%%%%%%%%%%%%%%%%%%%%%%%%%%%%%%%%%%%%%%%%%%%%%%%%%%%%%%%%%%%%%%%%%
%  
%

\newpage

\newpage

\newpage

\newpage

\newpage

\newpage

\newpage

\newpage

%%%%%%%%%%%%%%%%%%%%%%%%%%%%%%%%%%%%%%%%%%%%%%%%%%%%%%%%%%%%%%%%%%%%%%%
%%%%%%%%%%%%%% End of Figures
%%%%%%%%%%%%%%%%%%%%%%%%%%%%%%%%%%%%%%%%%%%%%%%%%%%%%%%%%%%%%%%%%%%%%%%


\begin{references}
\bibitem{x-section} We have used the fact that at high energies the cross 
sections  scale as : 
$\sigma_{\nu_\tau}(E) = 0.67 \times 10^{-38} E$ cm$^2$/GeV and 
$\sigma_{\bar \nu_\tau}(E) = 0.34 \times 10^{-38} E$ cm$^2$/GeV.
\end{references}
\end{document}